\documentclass[pre,showpacs,onecolumn,preprintnumbers,superscriptaddress,amsmath,amssymb,color,amsfonts]{revtex4-1}
\usepackage{graphicx,subfigure}
\usepackage{epstopdf}
\bibstyle{apsrev4-1}
\usepackage{color}

\usepackage{mathrsfs}

\newsavebox{\astrutbox}
\sbox{\astrutbox}{\rule[-5pt]{0pt}{20pt}}

\begin{document}
\title{Impact of hydraulic tortuosity on micro/nanoporous flow}
\author{Shiwani Singh}
\email{singh.shiwani@gmail.com}
 \affiliation{Mathematics Institute, University of Warwick, Coventry CV4 7AL, United Kingdom}

 \begin{abstract}
Using the porous structures made up of homogeneously arranged solid obstacles, we examine the effects of rarefaction on the hydraulic tortuosity in the slip and early transition flow regimes via extended lattice Boltzmann method. We observed that modification in either the obstacle's arrangement or the porosity led to a power-law relation between the porosity-tortuosity.
% 
% First, a power-law relation between the porosity-tortuosity is observed by either modifying the obstacle's arrangement or the porosity.
Along with this, we also found that in the slip flow regime, the exponent of this relation contains the effect of finite Knudsen number (Kn). In addition, we  observed that on properly scaling Kn with porosity and hydraulic tortuosity, a generalized correlation can be obtained for apparent permeability.
%using the right scaling of Kn with porosity and hydraulic tortuosity, a generalized correlation can be obtained for apparent permeability.
\end{abstract}
\maketitle
\section{Introduction}

 Numerous scientific and engineering applications, including  water percolating through  soils,  gas transportation,  CO$_2$ sequestration, oil extraction with or without polymer flooding  and a host of others, can benefit from an understanding of the transportation mechanisms and fluid flow across  porous media \cite{bear1972dynamics,ghanbarian2013tortuosity,metz2005ipcc,lake1989enhanced}.  Porous media are made up of a solid matrix of material that is filled by a network of void spaces (pores) containing fluid and are connected by throats that are significantly smaller in size. One of the main physical properties of interest  is the permeability of a porous material which measures its ability to allow fluids (gas or liquid) to flow through it. Apart from porosity which is defined as the percentage of a porous sample that is occupied by pore space,  permeability also depends on the geometry and structure of the pores \cite{carman1937fluid,bear1972dynamics,dullien2012porous}.
%  
%  defined as the fraction of porous sample occupied by pore space,

  Depending on the geometry  and the location of the pores, the actual path taken by the fluid can be very complicated or tortuous.  Therefore, a parameter, hydraulic tortuosity (T), was introduced to take care of complicated transport paths in a comprehensive manner \cite{bear1972dynamics,koponen1996tortuous}. Hydraulic tortuosity can be understood as the ratio of the  average length of true flow routes to the  system's length in the direction of the macroscopic flow. 
 The optimal way to calculate this ratio would be to take the mean flow length from a weighted average of the streamline, however if the geometry is too complicated, it would not be feasible \cite{matyka2008tortuosity,ghanbarian2013tortuosity}.
%  Average flow length obtained from a some-sort of  weighted average of the streamline would be the ideal method for evaluating this ratio, but it might be impractical if the geometry is overly complex \cite{matyka2008tortuosity,ghanbarian2013tortuosity}. 
 To overcome this limitation, Koponen et al. \cite{koponen1996tortuous} and Duda et al. \cite{duda2011hydraulic} proposed a straightforward formula for T  as a ratio between the mean of total fluid velocity and the mean of component of the  velocity along the external force direction, which is extremely helpful in the situation of arbitrary geometry.
%   {\color{red} Ideally,  this ratio can be evaluated by doing some sort of weighted average of streamline which might prove to be useless if the geometry is too complicated \cite{}. However, in the Refs \cite{}, a simple way of calculating  T as a ratio of the mean fluid velocity
% to the mean component of the fluid velocity along the external force direction which is very useful in the case of arbitary geometry.
%   
% %   to evaluate actual path... .....
% %  to propose a universal, efficient
% % method of calculating hydraulic tortuosity in an arbitrary
% geometry
% }
 
%  {\color{blue} 
 Furthermore, the emergence of unconventional energy sources, like ultra-tight gas reservoirs within shale rocks, 
 have shown great potential  towards mitigating the world energy crisis \cite{javadpour2007nanoscale,freeman2011numerical}.
  Shale rocks  are highly tortuous  and are made up of fine-grained material which contains pores in the nanoscale size range. At this scale where the mean free path of gas molecules becomes equal to or greater than the characteristic flow length within ultra-tight rocks, the rarefaction effects starts to emerge.  The Knudsen number Kn, or the ratio of the mean free route of gas molecules to the typical flow length, indicates the degree of rarefaction.  The fluid behavior can be separated into four primary categories based on the Kn value: continuum flow regime with Kn $<$0.001, slip flow regime with $0.001< {\rm Kn} < 0.1$, transition flow regime with $0.1 < {\rm Kn} < 10$, and free molecular flow regime with Kn $> 10$. 
 Contrary to what is predicted by Darcy's law in continuum flow regime, rarefaction effect cause  gas permeability (apparent) to increase as the pore size decreases. 
 Kinkenberg was the first to claim that this increase is caused due to the rise in gas slippage at the solid-fluid interface \cite{klinkenberg1941permeability}. Therefore, to precisely predict the reservoir's production capacity and longevity, it becomes unquestionably crucial to investigate the effect of rarefaction on physical parameters like permeability and tortuosity of ultra-tight porous media.%  
%  it becomes unequivocally important to study the effect of rarefaction on physical properties like permeability and tortuosity of porous media to accurately estimate the gas production and lifespan of the reservoir. 
%  
%  the lenght rarefaction effect becomes important
 
%  }

   The goal of this research is to provide a deeper understanding of the gas transport characteristics and Kn dependency of various physical properties of porous media, including tortuosity, in carefully designed porous media.
  The set-up is designed in a simple way where the circular obstacles are arranged homogeneously between two parallel plates. The porosity is varied by changing the diameter of the obstacle and the tortuosity is altered by changing the location of next nearest obstacle.
   As a simulation technique, we used the lattice Boltzmann (LB) method which has not only proven to be a useful tool for simulating Newtonian continuum hydrodynamics \cite{benzi1992lattice,chen1998lattice,succi2001lattice,aidun2010lattice,kruger2017lattice} but also been successfully extended for flows beyond Navier-Stokes in the past few years particularly to non-equilibrium ( finite Kn flows). The extended LB method  utilizes either a  regularization procedure \cite{zhou2006simulation,zhang2006efficient,latt2006lattice} or an appropriate multi-relaxation time model \cite{tao2015boundary,guo2014generalized} in combination with kinetic boundary condition \cite{ansumali2002kinetic,montessori2015lattice,singh2017impact,singh2017influence}. We also conducted a thorough parametric study with the goal of determining a consistent way to account for the impact of porosity and tortuosity on gas permeability.

 Starting from a brief overview of tortuosity and its  evaluation technique in Section \ref{tor}, the rest of the paper is organized as follow: In Section \ref{rpm}, the representative porous media set-up is detailed and a reference has been made to simulation technique, the extended lattice Boltzmann method, which is  detailed later in the Appendix \ref{LBM}. This followed by the Kn depended flow investigation in Section \ref{result} where starting by investigating  the local flow profile at various physical parameter like porosity and porous arrangement, we studied the effect of Kn on tortuosity-porosity relation in Section \ref{KnonT}. Further, the effect of tortuosity and porosity on gas permeability was studied in Section \ref{KnK}. Finally, the work is summarized and some future aspects of the work are discussed in Section \ref{conclude}.

\section{\label{tor} Tortuosity }
Hydraulic tortuosity (T) is the ratio of elongation of fluid streamlines due to the presence of obstacle (porous media) to the system size in the case of free flow \cite{matyka2008tortuosity,ghanbarian2013tortuosity}. Therefore, if $\lambda$ is the mean distance covered by fluid element and L is the system size in the direction of flow, the hydraulic tortuosity is defined as
\begin{equation}
 T=\frac{\lambda}{L}.
 \label{tr}
\end{equation}
By this definition, hydraulic tortuosity is always greater than or equal to 1 $(T\geq1)$. It means in a plane channel flow, $T=1$ since streamlines face no hindrance in the absence of of porous material. The value of $T$ rises when tortuosity grows because fluid has to travel farther through porous media.
   Since T is defined as deviation  of fluid path, it can be calculated using velocity field. Some of the methods used to calculate $T$ focuses on the  calculation of weighted averages of discrete streamline \cite{knackstedt1994direct,zhang1995direct}. Most of these methods will be difficult to apply in complex geometries. 
   However, in the works of   Koponen et al. \cite{koponen1996tortuous} and Duda et al. \cite{duda2011hydraulic},  the authors came up with a simple approach for the  incompressible and non-reentrant flow to calculates $T$  directly from the velocity field by averaging its components  in the following manner:
   \begin{equation}
    T=\frac{\langle u\rangle}{\langle u_x\rangle}
    \label{tort}
   \end{equation}
   where $\langle u\rangle$  is the average of the magnitude of fluid velocity and $\langle u_x\rangle$ is the average of its component along the direction of flow. Hereafter, in the present study, we will be using the above mentioned definition (Eq. \eqref{tort}) which  is now being routinely used to define $T$ \cite{icardi2014pore,matyka2016power,muljadi2016impact,el2019pore,graczyk2020predicting}. 
   A detailed pedagogical review about the approach used in the Ref.\cite{duda2011hydraulic} can be found in the Ref. \cite{matyka2012calculate}.

%    To understand there approach in detail apart from the Ref.\cite{duda2011hydraulic}, a pedagogical review of the same is also provided by the authors in Ref. \cite{matyka2012calculate}.

 \section{ \label {rpm} Physical porous media}
\begin{figure}[h]
\subfigure[$\theta_0$, $\phi=0.75$]{
   \includegraphics[scale=0.2]{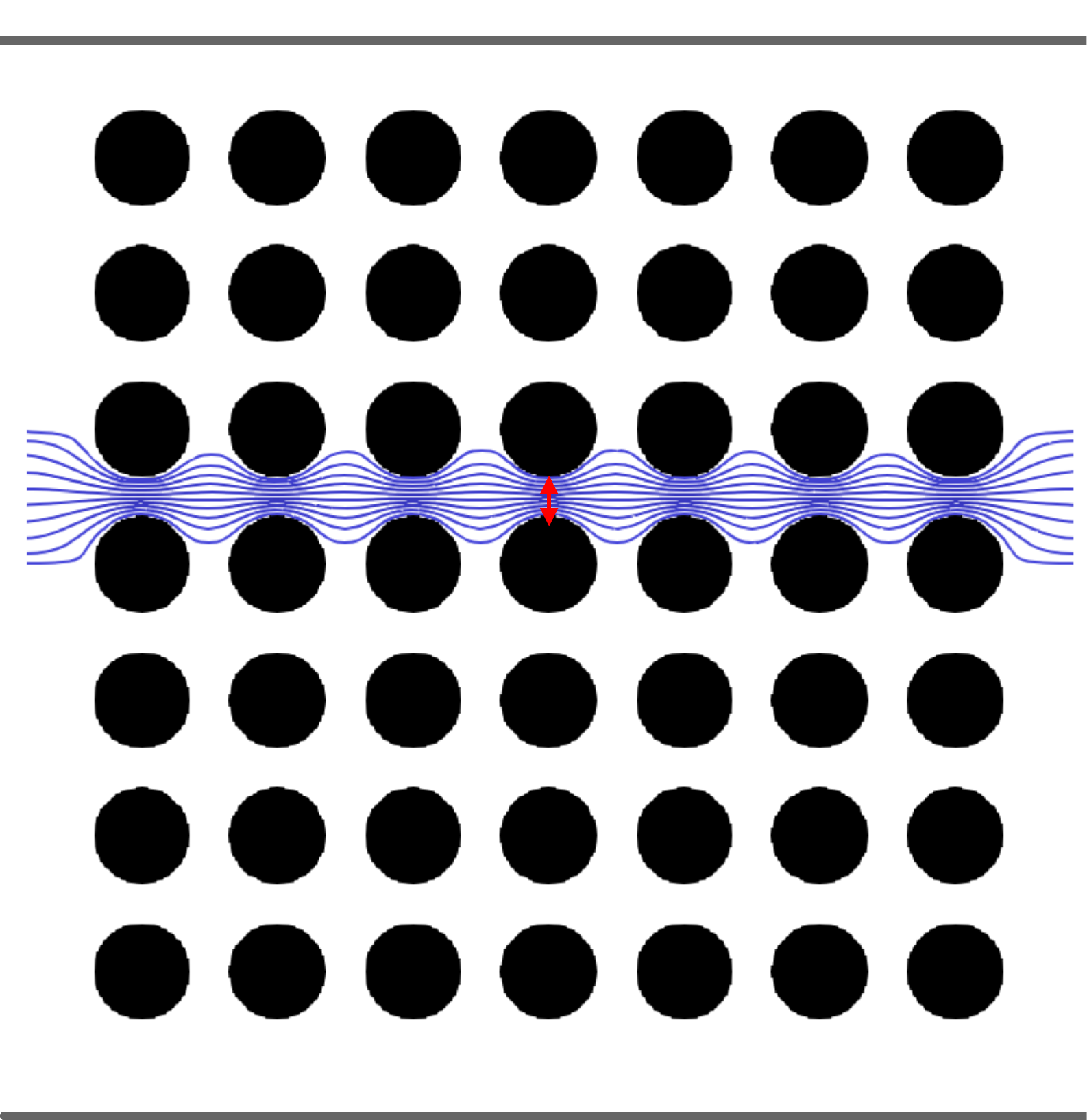}\label{het0_phi075}}
\subfigure[$\theta_6$, $\phi=0.75$]{
\includegraphics[scale=0.2]{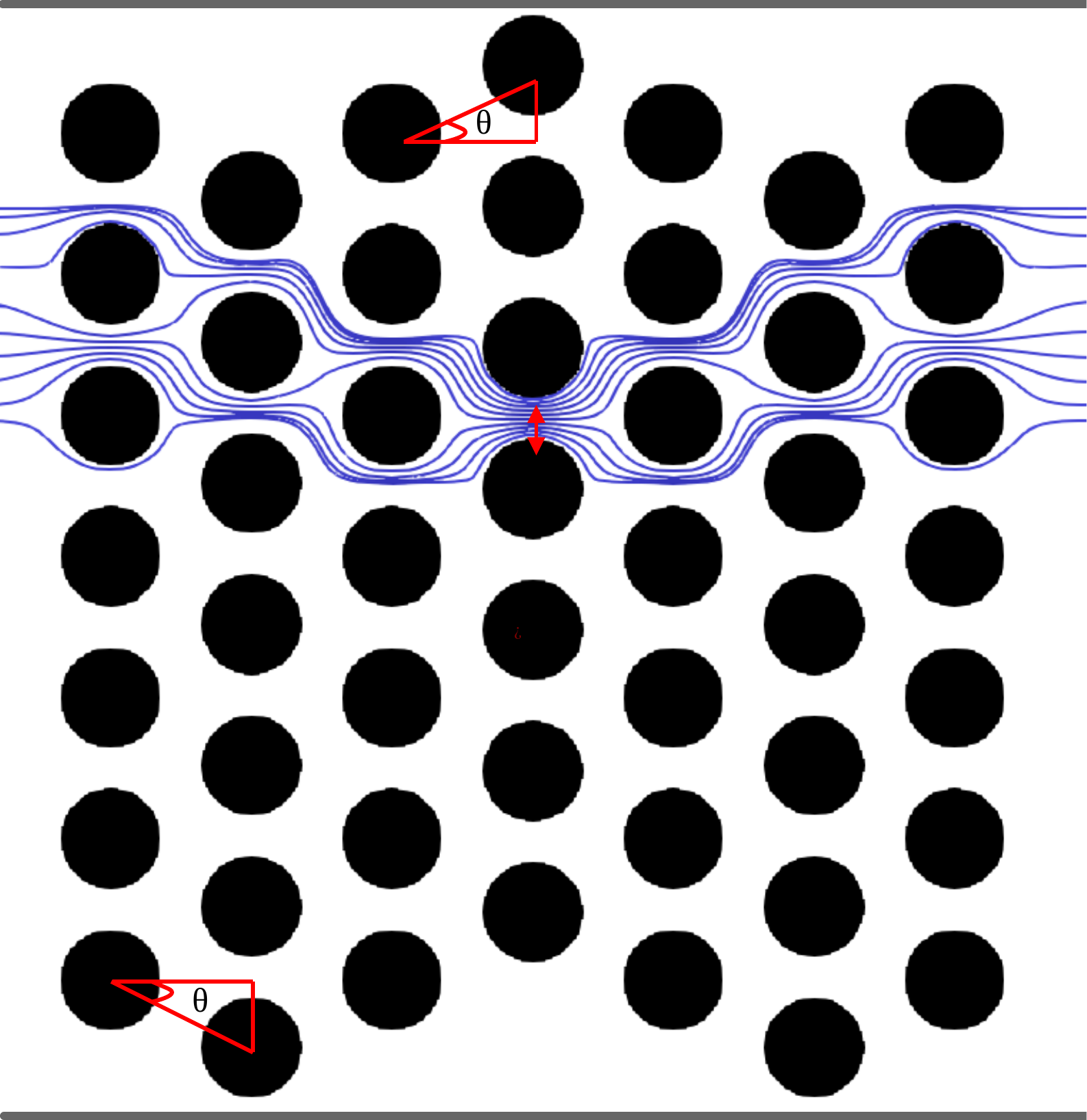}\label{het6_phi075}}
\subfigure[$\theta_6$, $\phi=0.90$]{
\includegraphics[scale=0.2]{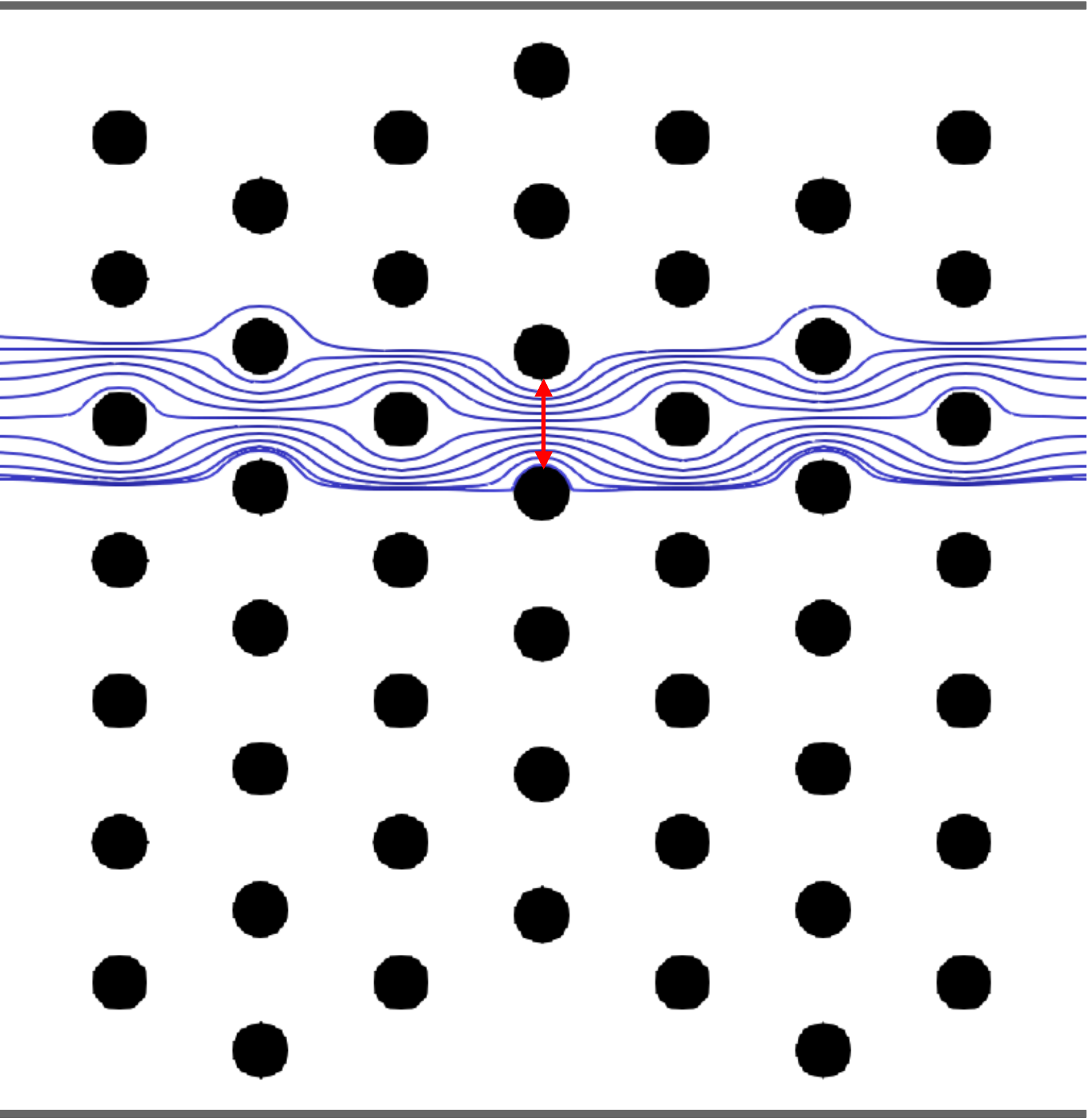}\label{het6_phi09}}
\caption{Schematic of representative porous media designed as an array of circular obstacle in two-dimension}
 \label{por_rep}
  \end{figure}

 In the present study, we used  simple two-dimensional homogeneous geometries constructed by placing circular obstacle in seven rows and seven column  between two parallel plates to study the effect of tortuosity (see Fig. \ref{por_rep}). The lattice Boltzmann (LB) method  is used to solve the flow equations.  In the recent past, the  LB method has emerged as an powerful tool to solve the continuum hydrodynamics \cite{benzi1992lattice,succi2001lattice}  and with some physics inspired refinements has shown to capture fluids dynamics beyond Navier-Stokes equations\cite{montessori2015lattice,singh2017impact,mohammed2021lattice}. The details of this refined method is presented in the Appendix \ref{LBM}. 
 
%  In the context of the LB method, Kn number is defined as $.$ 
%  η/(h cs ),
% where the kinematic viscosity is defined as η = c2
% s (τ − t/2)
In the aforementioned geometrical set-up, the lattice node which falls on or inside solid is marked as `$\rm node_{S}$' and the ones occupied by fluid are marked  $`\rm node_{F}$'. Hence, the calculation of the porosity( $\phi$), defined as the volume (area in two-dimension) of void to the total volume (area), is straight forward and is given as
 \begin{equation}
  \phi=\frac{\rm node_{F}}{\rm node_{F}+\rm node_{S}}.
 \end{equation}
 At this point, it is worth introducing ${\rm Kn}$ which is the
 ratio of the molecular mean free path with respect to character macroscopic length. The Kn can also be represented in terms of  kinematic viscosity, $\eta$, as ${\rm Kn}=\eta/(d c_s)$  where we chose $d$  to be the smallest pore-throat diameter present as red double arrow in the Fig. \ref{por_rep},  and $c_s$ is the speed of sound which is explained later in the Appendix \ref{LBM} where LB method is elaborated alongside with regularization mechanism which filters out the non-hydrodynamic moments and the kinetic boundary condition (KBC) which is based on diffusively reflecting wall. In the recent past, it has been observed that regularization and KBC are  the crucial ingredients required in the standard LB for the simulation of finite Kn flow   \cite{montessori2015lattice,singh2017impact}. The simulations were performed with 500 grid points in each direction which resulted in 24-38 number grid points representing each pore throat. 
 
%  Furthermore, the fluid behavior can be separated into four primary categories based on the Kn value: continuum flow regime with Kn $<$0.001, slip flow regime with $0.001< {\rm Kn} < 0.1$, transition flow regime with $0.1 < {\rm Kn} < 10$, and free molecular flow regime with Kn $> 10$.
%  
%  The Kn can be used to define the rarefaction of the flow and fluid behaviour can be divided in four major: continuum flow regime with
% Kn $<$ 0.001, slip flow regime with 0.001 $< {\rm Kn} < 0.1$, transition flow
% regime with $0.1 < {\rm Kn} < 10$, and free molecular flow regime
% with Kn $> 10$.
%}

  Before studying the Kn dependent rarefaction effects, we focused in the continuum regime to establish the relation between the porosity and tortuosity. Consequently, starting from uniform geometry where distance between the next nearest circle placed in row is  the same  as that of the one placed in column (Fig. \ref{het0_phi075}),  the simplest way to alter tortuosity is to change the arrangement  by placing  the next nearest circle at a distance defined by the angle $\theta$ as shown in Fig. \ref{het6_phi075}. This allowed us to do a controlled study and have insight of how tortuosity increases/decreases with the arrangement of obstacles in the media. 
      \begin{figure}[h]
   \includegraphics[scale=0.4]{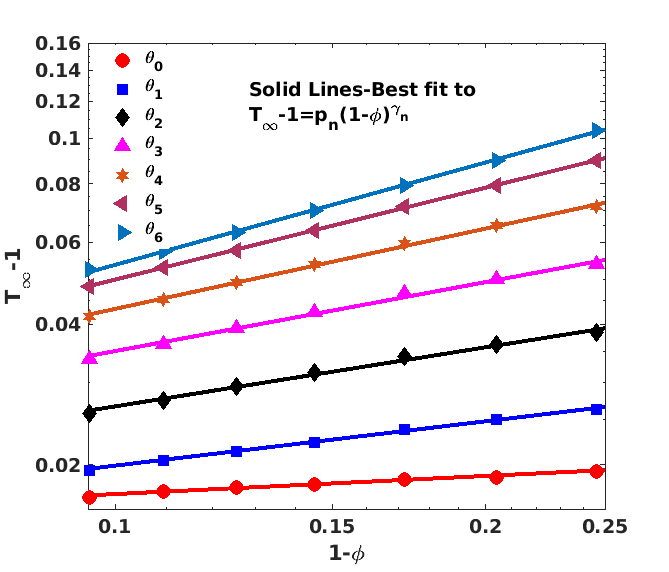}
\caption{Tortuosity as a function of porosity for various arrangement in porous media. Solid lines are the best  fit to $T-1=p_n(1-\phi)^{\gamma_n}$ where $n (0,1,...,6)$ defines the alignment.}
   \label{tor_por}
  \end{figure}
 We choose the following values of alignments: $[\theta_0,\theta_1,\theta_2,\theta_3,\theta_4,\theta_5,\theta_6]=[0,5.5^\circ,9.162^\circ,10.45^\circ,14.47^\circ,17.88^\circ,21.96^\circ,25.82^\circ]$.
 The streamlines  passing through the pore throat, present nearly at the middle of the 
 domain,  marked with a red double arrow shows that keeping the porosity same, tortuosity appears to increase with $\theta$ (see Fig. \ref{het0_phi075} and Fig. \ref{het6_phi075}). In Fig. \ref{het6_phi09}, we kept the alignment ($\theta$) same as that of Fig. \ref{het6_phi075} but increased the porosity which resulted in an obvious decrease in tortuosity. Recall, as defined earlier, larger the average length of streamlines, larger will be the tortuosity.
To calculate the absolute tortuosity, $T_\infty$,
 for each porous configuration in the limit of Kn $\to$ 0, we
extrapolated the tortuosity (T) calculated within the range from
${\rm Kn} = 10^{-1}$ to ${\rm Kn} = 10^{-3}$ and chose the value at ${\rm Kn}=10^{-7}$ as  $T_\infty$.

  Fig.  \ref{tor_por} represents the  the absolute tortuosity ($T_\infty$) as function of porosity for large to medium porosity  ranging from 0.90 to 0.75 for all the seven alignment ($\theta_0 \textrm{ to }\theta_6$).  Firstly, the figure shows  the obvious trend that  with increase in porosity, tortuosity decrease for every configuration. Secondly, the log-log plot between $T_\infty-1$ and $1-\phi$ clearly shows a power-law behaviour  with the  exponent  as   
  $[\gamma^0, \gamma^1, \gamma^2, \gamma^3, \gamma^4, \gamma^5, \gamma^6]=\left[0.1286, 0.3153, 0.4199, 0.4902, 0.5706,0.6530, 0.7279\right]$ and constants   as $[p_0,p_1, p_2, p_3, p_4, p_5, p_6]=\left[0.0233, 0.0412, 0.0702, 0.1084, 0.1606,0.2243, 0.2866\right]$ (see Fig. \ref{tor_por}). This is an empirical relationship, however, is  in agreement with the two-dimensional flow with randomly distributed square obstacle as used by Duda et. al \cite{duda2011hydraulic} for large porosity ($<0.8$) which also showed a power-law behaviour as $T-1\sim (1-\phi)^\gamma$, however their  exponent ($\gamma$) was $1/2$.

\section {\label{result} Flow at finite Knudsen Number}
  %%%%%%%
  \begin{figure*}[h]
   \includegraphics[height=100mm,width=150mm]{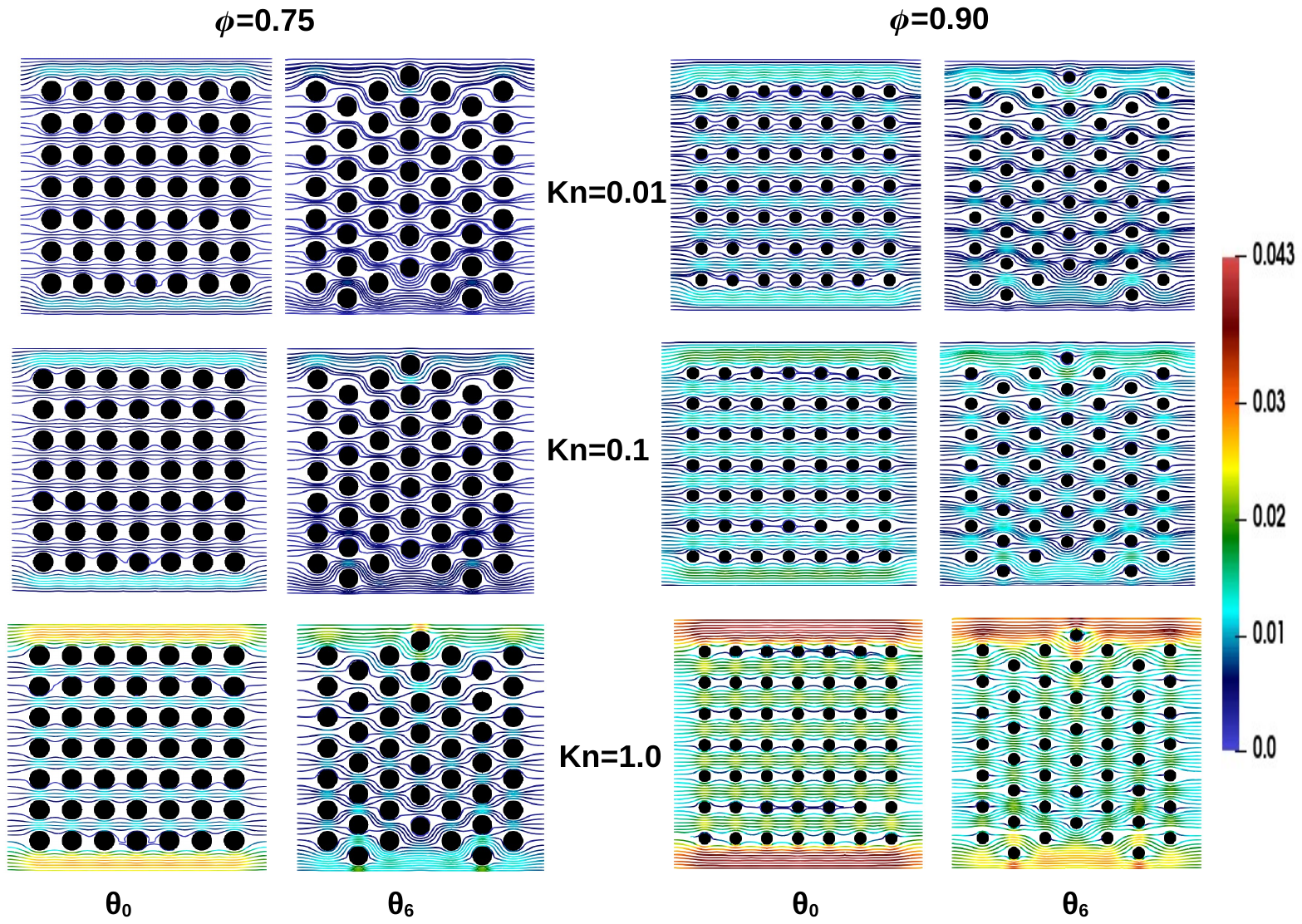}
   \caption{Steady state velocity streamlines for  porosity, $\phi=0.75$ and $\phi=0.90$, porous arrangement , $\theta_0$ and $\theta_6$, and at Kn=0.01, Kn=0.1 and Kn=1.0.}
   \label{flow_field}
  \end{figure*}
   Tortuosity and permeability are two important properties that influence the passage of fluid through a porous media. In the subsequent sections, we will investigate the effect of rarefaction on these two properties. However, firstly, it would also be interesting to observe the behavior of the local velocity in different porous arrangement. Therefore, in Fig. \ref{flow_field}, we plotted steady state streamlines at different Kn for  porosities, $\phi=0.75$ and $\phi=0.90$, and at pore alignments, $\theta_0$ and $\theta_6$. The scale for the magnitude of streamlines in all the cases has been kept the same to show a relative difference between all the considered arrangements. From the Fig. \ref{flow_field}, we can make the following observations:
  \begin{itemize}
%   \end{enumerate}
% 
%   
\item For a given porosity ($\phi$), the flow take more tortuous path as the uniformity of the obstacles decreases (i.e. from $\theta_0$ to $\theta_6$) at every Kn.
\item In the current configuration, when porosity rises, the pore throat widens, facilitating fluid flow across the porous medium. As a result, the velocity inside the pore-throat increased as the $\phi$ value increased.
   \item However, due to Knudsen diffusion, there is a non-zero fluid velocity at solid barriers, which causes the velocity inside the pore throat to grow with increasing Kn in all situations.
%    
%    Due to the Knudsen diffusion, there exists a non-zero fluid velocity at the solid boundaries which results in the  increases of the magnitude of velocity  inside the pore throat with increasing Kn in all the cases.
%    \item In the present set-up, as the porosity increases, the pore throat becomes big which facilitates  the fluid movement through the porous medium. This resulted in a increase in velocity inside the pore-throat with increasing $\phi$.
  \end{itemize}

% We see from the figure that at small Kn (0.01) which is close to continuum flow, the streamlines at more ....
% Due to Knudsen diffusion, there exists a non-zero fluid velocity which results in the  increases of the magnitude of velocity  inside the pore throat with increasing Kn. 

  %\section {\label{result}Result}

% \begin{figure}[h]
%    \includegraphics[scale=0.4]{figures/perm_por.png}
% \label{cmb}
% \caption{write}
%   \end{figure}
\subsection{\label{KnonT} Effect of finite Kn on tortuosity}
 \begin{figure*}[h]
% \subfigure[homo]{
%    \includegraphics[scale=0.3]{figures/homo_kn_new.png}}
\subfigure[$\theta_1$\label{ht1}]{
\includegraphics[scale=0.3]{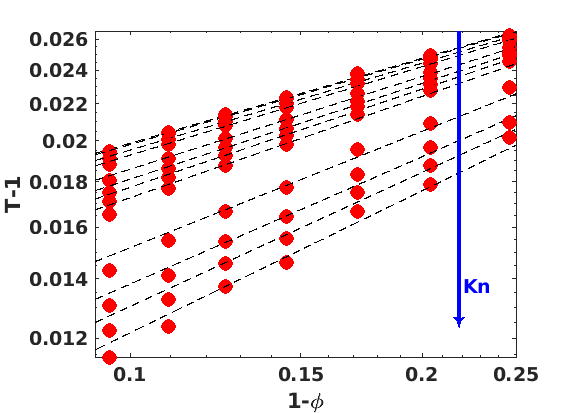}}
%\subfigure[hetro-2]{
%\includegraphics[scale=0.3]{figures/het2_kn_new.png}}
\subfigure[$\theta_3$\label{ht3}]{
\includegraphics[scale=0.3]{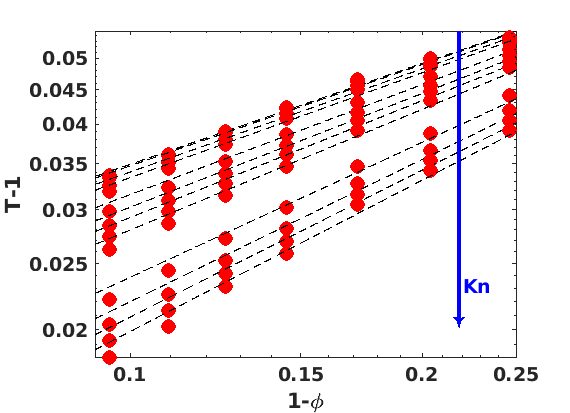}}
%\subfigure[hetro-4]{
%\includegraphics[scale=0.3]{figures/het4_kn_new.png}}
%\subfigure[hetro-5]{
%\includegraphics[scale=0.3]{figures/het5_kn_new.png}}
\subfigure[$\theta_6$\label{ht6}]{
\includegraphics[scale=0.3]{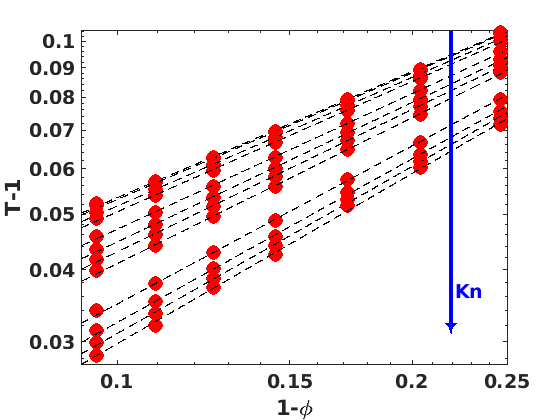}}
\caption{Tortuosity as a function porosity at pore alignment $\theta_1$, $\theta_3$ and $\theta_6$ with varying Kn. Dashed lines are the best fit to the function $T-1=p_n(1-\phi)^{(\gamma_n+{\rm f}_n({\rm Kn}))}$ where the value of $n$(0,1,2,..6) depends on the porous arrangement.  
%The value of $\gamma_n$ and $p_n$ are the same as evaluated in Section. \ref{rpm} (Fig. \ref{tor_por}).
}
\label{Kn_tor_fn}
  \end{figure*}
Hydraulic tortuosity (T) is the measure of average fluid streamline length in a porous medium versus system length of obstacle-free flow, and as defined in Eq. \ref{tr}, can be calculated using appropriate averages of velocity field component. This makes T a flow dependent observable. Therefore, any effects of rarefaction that appears on velocity field such as, increment of slip velocity with Kn, also have significant effect on T.

To investigate this effect, we  computed the value of T as a function of porosity, $\phi$, with different Kn for all the arrangement of porous medium, $\theta_0$-$\theta_6$ which again shows a power law behavior, however with a different exponent (Kn-dependent) as shown in Fig. \ref{Kn_tor_fn}.  When Kn rises, the slip velocity makes it easier for fluid to move through the pores, which causes a  concomitant decrease in the  hydraulic tortuosity. This behavior is clearly reflected in Fig. \ref{Kn_tor_fn}.
% The slip velocity makes it easier for fluid to move through the pores as Kn rises, and the hydraulic tortuosity is anticipated to decrease, as seen in Fig. \ref{Kn_tor_fn}.
In order to fully contain the Kn-dependency of hydraulic tortuosity in the exponent of the power-law behavior, we plotted the $T-1$ and $1-\phi$ with the best fitted lines to the function $ T-1=p_n(1-\phi)^{(\gamma_n+{\rm f}_n({\rm Kn}))}$ in Fig. \ref{Kn_tor_fn} where $n$ (0,1,2,..6)  dictates the pore arrangement.  Here, the values of $\gamma_n$ and $p_n$ are the same to those assessed in Section \ref{rpm} (Fig. \ref{tor_por}). 
%{\color{blue} 
% If the tortuosity-porosity relation is represented as $ T-1=p_n(1-\phi)^{(\gamma_n+{\rm f}_n({\rm Kn}))}$ where $n$ (0,1,2,..6)  dictates the pore arrangement, the dependence of this exponent on Kn becomes clear.  Here, the values of $\gamma_n$ and $p_n$ are the same to those assessed in Section \ref{rpm} (Fig. \ref{tor_por}). The dependence on the Kn number is now contained in $f_n$, which is the main reason behind writing $T-1$ in this manner.
% Keeping all the configuration same and changing Kn reveals that, as Kn increases, the exponent of power law also increases which makes ${\rm f}_n$ an increasing function of Kn.
% It can also be observed from Fig. \ref{Kn_tor_fn} that as the non-uniformity increases from Fig. \ref{ht1} to Fig.\ref{ht3} and finally to Fig.\ref{ht6} by increasing the non-uniformity  of pores the from $\theta_1$ to $\theta_3$  and to $\theta_6$  respectively, the exponent in the early-transition regime (bottom few lines in every figure) appears to be less distributed with varying Kn. 
% % In other words, the impact of increasing Kn at high non-uniformity is reduced. This shows that in the early-transition phase, more non-uniformity reduces the influence of rarefaction. 
% In other words, there is less of an impact on physical properties when Kn is increased at high non-uniformity. This demonstrates that increased non-uniformity in the early transition phase lessens the impact of rarefaction. 
Changing Kn while maintaining the same configuration demonstrates that ${\rm f}_n$ is an increasing function of Kn because as Kn rises, the exponent of the power law rises as well. Fig. \ref{Kn_tor_fn} also shows that the exponent in the early-transition phase (bottom few lines in each figure) appears to change less profoundly with changing Kn as the non-uniformity between the pores increases from Fig.  \ref{ht1} ($\theta_1$) to Fig. \ref{ht3} ($\theta_3$) and eventually to Fig.  \ref{ht6} ($\theta_6$). This shows that when Kn is increased at high non-uniformity, there is less of an effect on physical attributes. In other words, the effect of rarefaction in early transition regime is diminished by higher non-uniformity.
But this can also be a result of a technical problem  associated with the method. It has been shown in earlier research that a lattice Boltzmann model with regularization and a diffuse wall kinetic boundary condition (as used in the current study) performs admirably well in the slip flow regime and near to the early-transition flow flow. However, to take into account the flow beyond this limit, one must utilize a higher order lattice, according to Ref\cite{montessori2015lattice}. This aspect of scheme needs further investigation, thus it is left for future study.
% 
% 
% 
% 
% It can also be observed from Fig. \ref{Kn_tor_fn} that as the non-unformity increase from Fig. \ref{ht1} to Fig. \ref{ht3} and finally to Fig. \ref{ht6} with increasing distribution of pores from
% $\theta_1$  to $\theta_3$ and to $\theta_6$   respectivelty, the exponent in the early-transition regime (bottom for lines in every figure), the increase in exponent is not....  The first reason behind this might be more physical since .....because due to construct of present setup, the increase of velocity at the wall in finite Kn flow will not propogate and redistribute for set up close to uniform () case to those which are highly non-uniform (). This will reduce overall increase of the velocity in .... direction for non-uniorm case, hence the value of hydraulic tortusity. The second reason might be more tecnical. As mention in previous studies \cite{}, a lattice Boltzmann in combination with regularization and diffuse wall kinetic boundary condition (as used in the present study) performs very well in the slip flow region and close to early-transition regime. 
% However, acccording to Ref. \cite{}, in order to consider the flow beyond this limit, one must use a higher order lattice. This aspect of the scheme require further investigation and is left for future study.
 \begin{figure*}[h]
   \includegraphics[height=75mm, width=150mm]{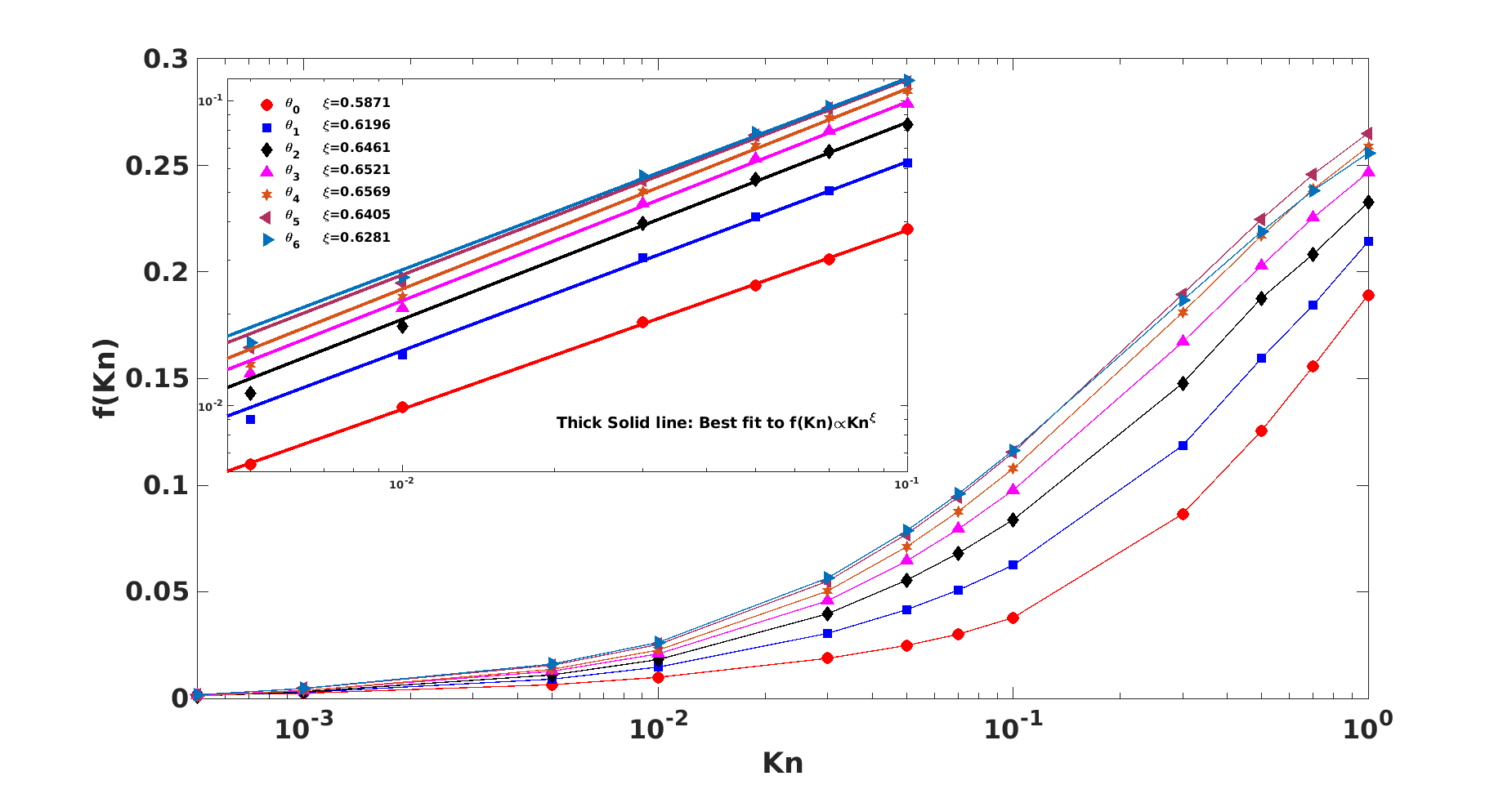}
\caption{The Kn dependent addition in the exponent of power-law behavior, $f_n$, for various  Kn. Here $n$ (0,1,...,6) describes the alignment of porous arrangement. The inset magnifies the slip flow regime which shows that $f({\rm Kn})\sim {\rm Kn}^{0.6}$. }
\label{fn_kn}
  \end{figure*}

Furthermore, to investigate the behavior of ${\rm f}_n$ with respect to Kn,  we plotted  ${\rm f}_n$ as a function of Kn (see Fig. \ref{fn_kn}) which firstly indicate that  ${\rm f}_n$ is indeed an increasing function of Kn for all the porous arrangements. However, in the slip regime ($0.001<{\rm Kn}<0.1$), ${\rm f}_n\sim {\rm Kn}^{0.6}$ (see the inset of Fig. \ref{fn_kn}). 
This suggests that there exists a consistent dependence of Kn on the tortuosity-porosity relation  at least in the slip flow regime  for the porous setup used here as, $ T-1\sim(1-\phi)^{(\gamma_n+{\rm Kn}^{0.6})}$.
% This suggest that atleast for the porous-setup chosen herein, there exist a constant dependence of Kn on tortuisity-porosity relation 

\subsection{\label{KnK}  Effect of finite Kn on permeability}  
  One of the  physical quantity of interest for the flow inside highly permeable porous media is the absolute permeability ($\kappa_\infty$) which is obtained by calculating the ﬂux ($Q$) at different pressure drops (or by adjusting the body force values) in the following manner:
   \begin{equation}
    \kappa_\infty=\frac{Q\eta}{\rho g}
   \end{equation}
where $\eta$ is the dynamic viscosity, g is the body force and $\rho$ is the density of the fluid. Permeability is a  crucial element in determining the transport capacity of porous media.  However, in unconventional reservoirs such a ultra-tight pores of shale rock, despite the presumption of absolute permeability being very low, experiments observed that apparent gas permeability (AGP) ($\kappa$), is much higher than  $\kappa_\infty$. Furthermore, the idea that the rise in gas slippage at the solid-fluid interface is to blame for the increased permeability was first put forward by  Kinkenberg who suggested the permeability correction factor (PCF), which is defined  as the ratio of apparent permeability ($\kappa$) to absolute permeability ($\kappa_\infty$), to be a linear function of Kn as:
\begin{equation}
\frac{\kappa}{\kappa_\infty}=1+4{\rm Kn}.
\label{klinkenberg}
\end{equation}
%{\color{blue}
Beskok and Karniadakis \cite{beskok1999report} further proposed  a second-order correlation that can  be used to describe all four fluid flow regimes and is given as:
\begin{equation}
\frac{\kappa}{\kappa_\infty}=[1+\alpha({\rm Kn}){\rm Kn}]\left(1+\frac{4{\rm Kn}}{1-b{\rm Kn}}\right)
\label{beskok_karniadakis}
\end{equation}
where slip coefﬁcient $b$ equals -1 for slip ﬂow, and $\alpha({\rm Kn})$ is the rarefaction coefﬁcient. The expression for $\alpha({\rm Kn})$ is somewhat complex, but Civan \cite{civan2010effective} later suggested one (Beskok \& Karniadakis-Civan’s correlation) that is considerably more straightforward:
\begin{equation}
\alpha({\rm Kn})=\frac{1.358}{1+0.170{\rm Kn}^{-0.4348}}.
\label{civan-slip_red}
\end{equation}
In addition, Civan \cite{civan2010effective} suggested that in the slip flow regime, $\alpha({\rm Kn})$ can also be neglected making PCF take the following form:
\begin{equation}
\frac{\kappa}{\kappa_\infty}=\left(1+\frac{4{\rm Kn}}{1+{\rm Kn}}\right).
\label{civan-slip}
\end{equation}
%}

\begin{figure*}[h]
% \subfigure[  $\theta_0$ ]{
%    \includegraphics[scale=0.125]{figures/het0.pdf}
%    \label{dbb_1_pt05}}
   \subfigure[  $\phi=0.75$ ]{
   \includegraphics[scale=0.19]{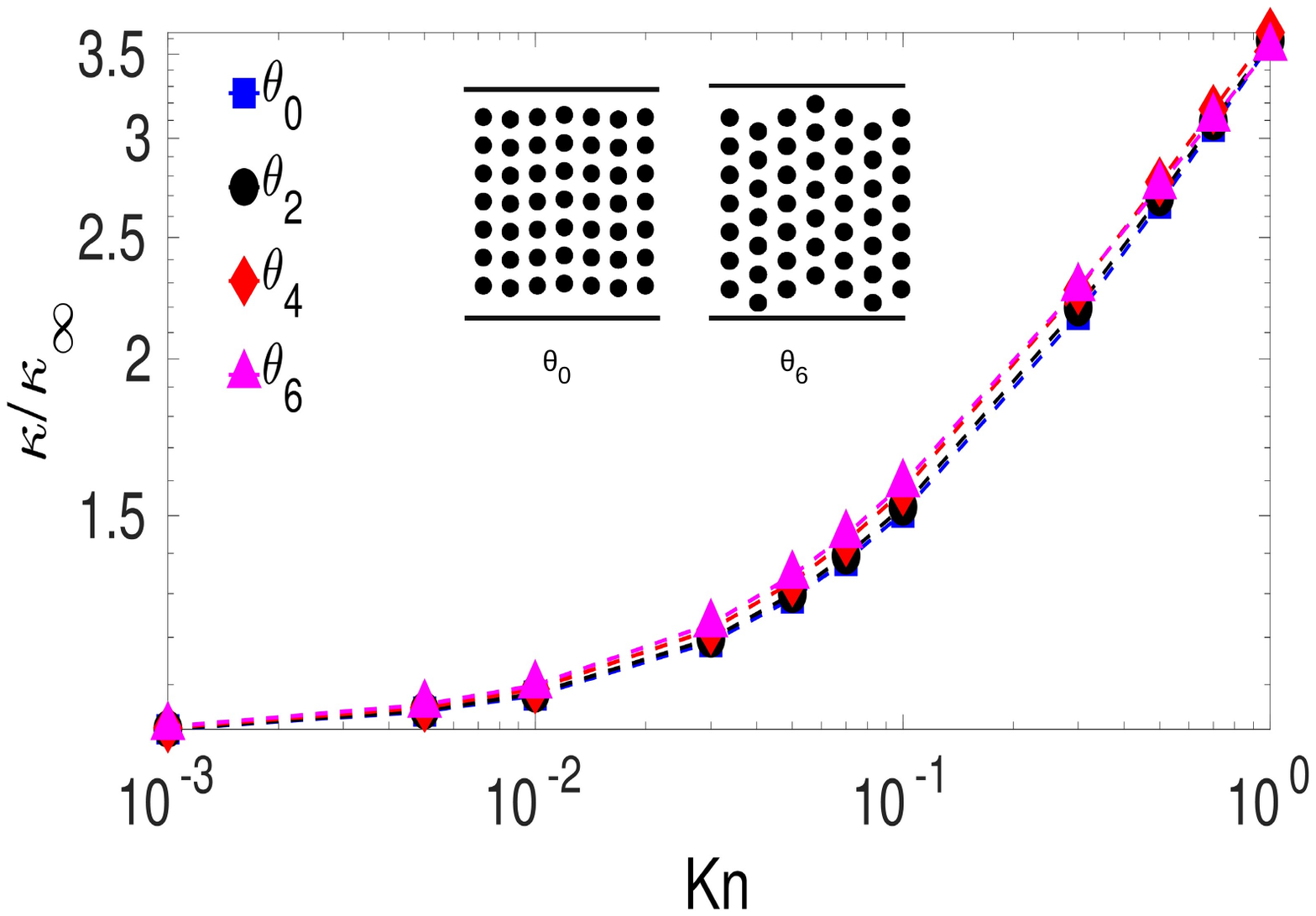}
   \label{fixpor_75}}
   \subfigure[  $\phi=0.83$ ]{
   \includegraphics[scale=0.19]{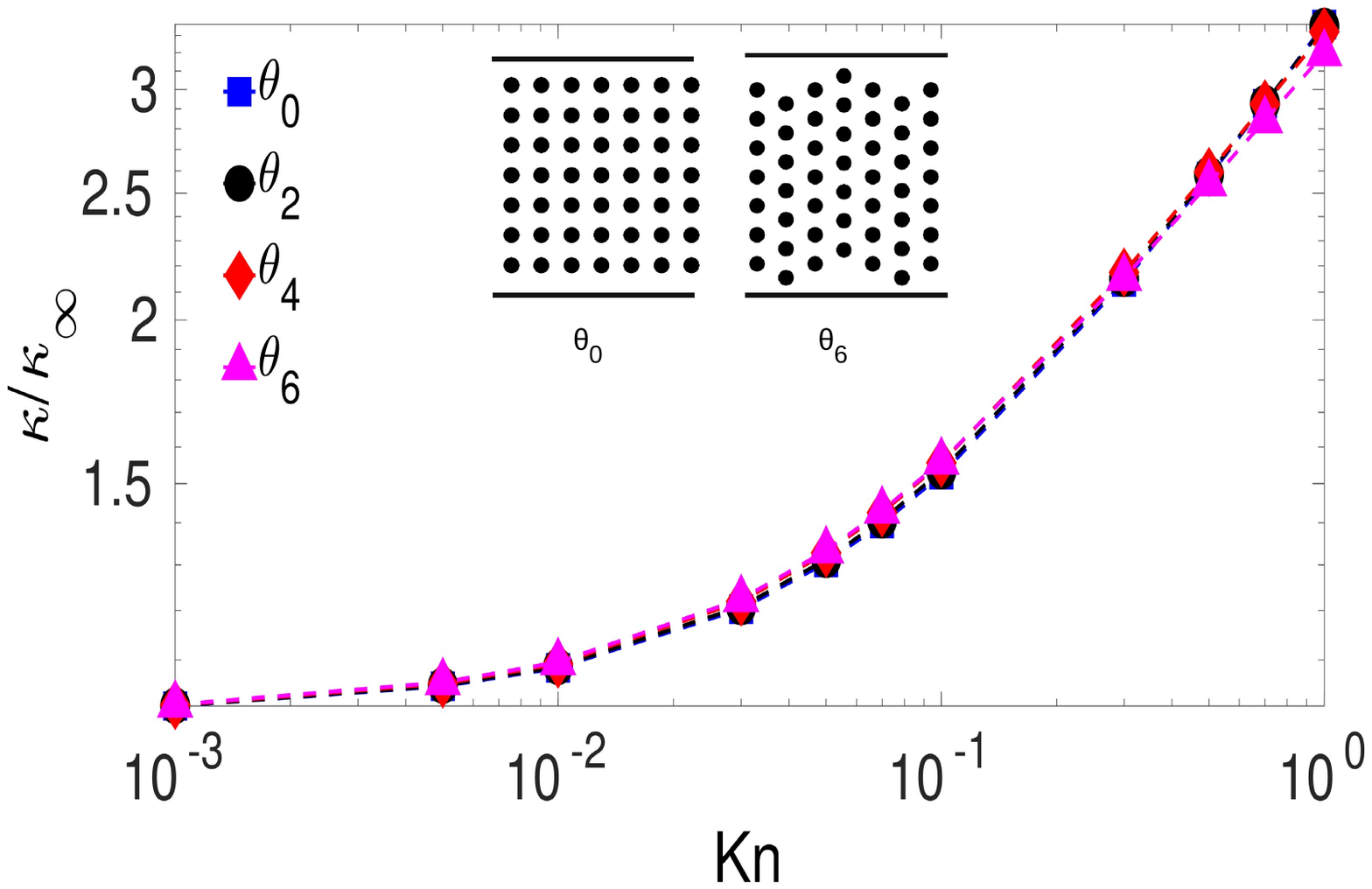}
   \label{fixpor_83}}
   \subfigure[ $\phi=0.90$ ]{
   \includegraphics[scale=0.19]{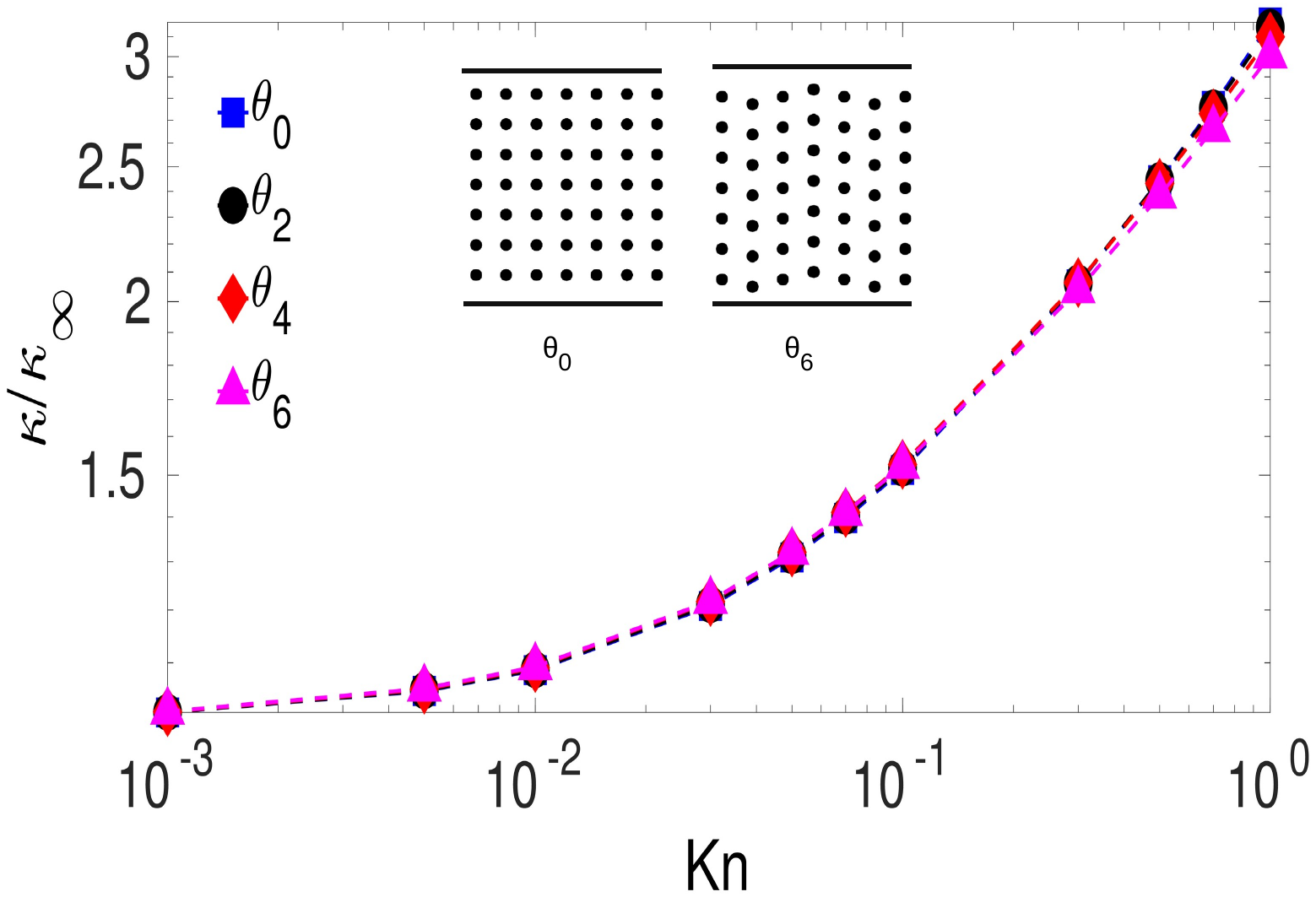}
   \label{fixpor_90}}
\caption{  The permeability correction factor (PCF) as a function of Kn at fixed porosity and varying alignments.
}
\label{perm_fixed_por}
  \end{figure*}

\begin{figure*}[h]
% \subfigure[  $\theta_0$ ]{
%    \includegraphics[scale=0.125]{figures/het0.pdf}
%    \label{dbb_1_pt05}}
   \subfigure[  $\theta_0$ ]{
   \includegraphics[scale=0.19]{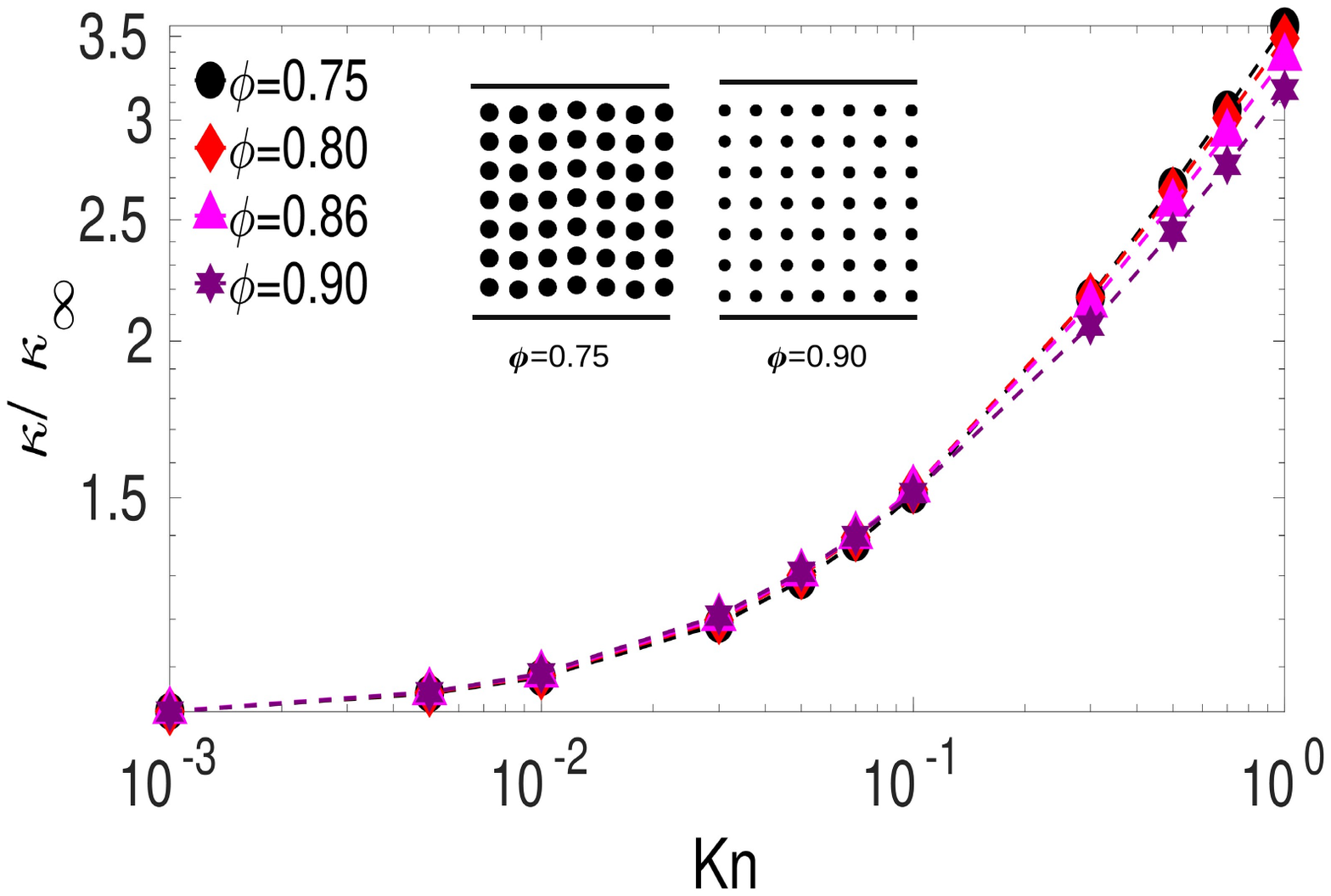}
   \label{fixalign_0}}
   \subfigure[  $\theta_3$ ]{
   \includegraphics[scale=0.19]{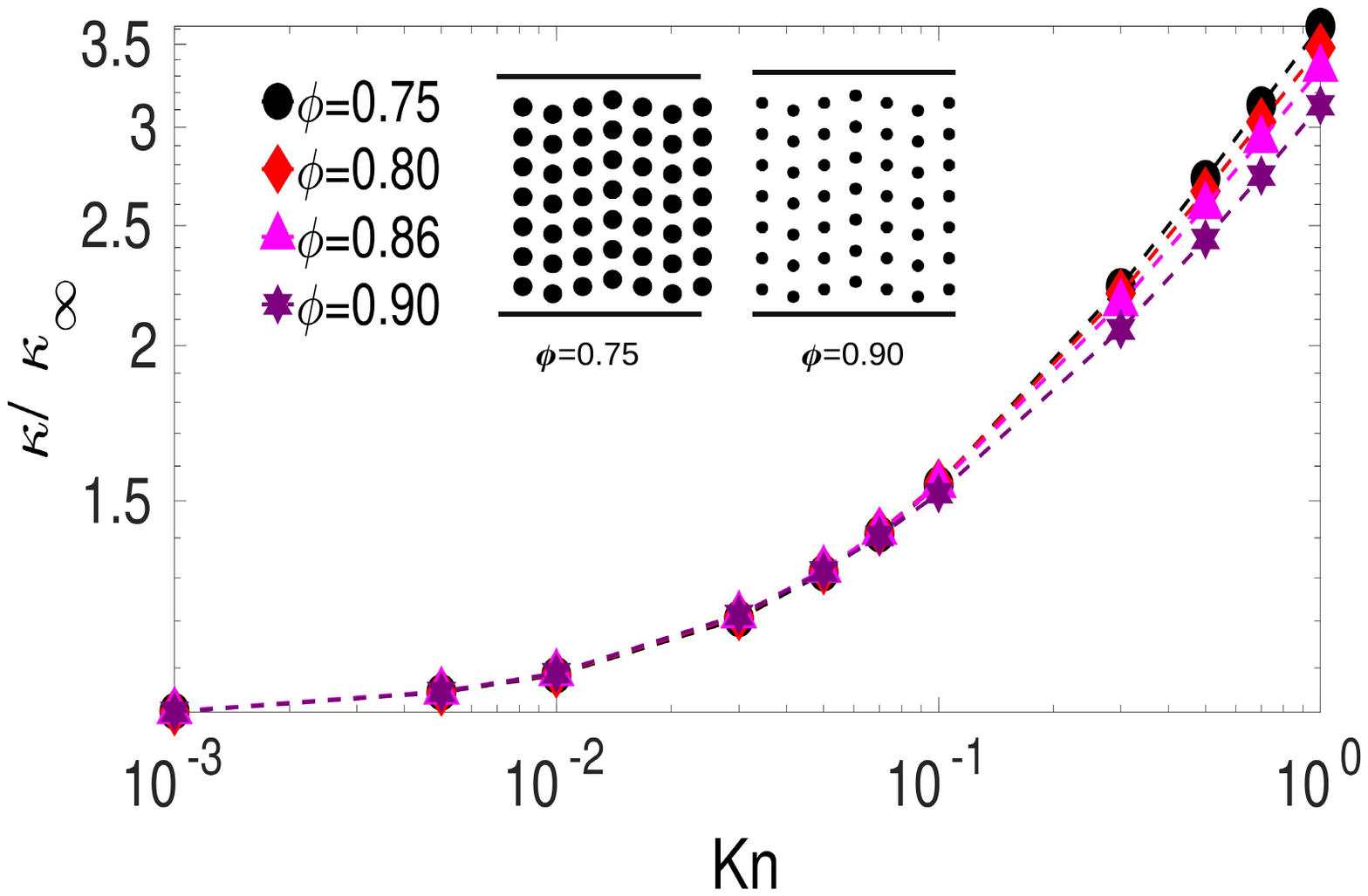}
   \label{fixalign_3}}
   \subfigure[ $\theta_6$ ]{
   \includegraphics[scale=0.19]{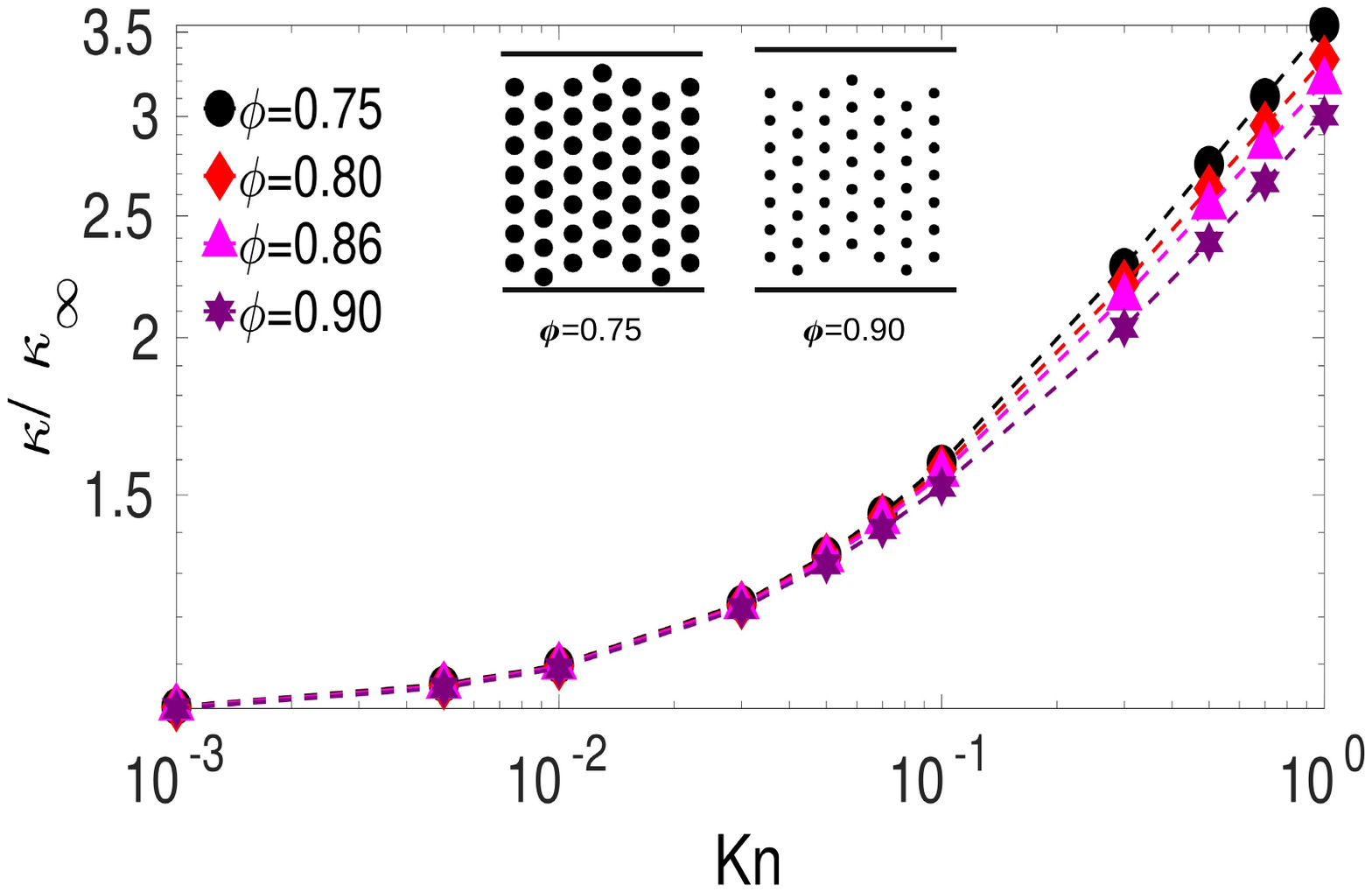}
   \label{fixalign_6}}
\caption{ The permeability correction factor (PCF) as a function of Kn at fixed alignment  and varying porosity.
% (Color online) Steady state velocity-streamlines with  diffuse bounce back boundary condition (DBB) with K obtained in three ways, K$_{1}$(Figs  \ref{dbb_1_pt05},\ref{dbb_1_pt5},\ref{dbb_1_5}),
% K$_{\rm norm}$(Figs  \ref{dbb_norm_pt05},\ref{dbb_norm_pt5},\ref{dbb_norm_5}) and K$_{\rm ad}$ (Figs  \ref{dbb_ad_pt05},\ref{dbb_ad_pt5},\ref{dbb_ad_5}),  for different Kn number
}
\label{perm_fixed_align}
  \end{figure*}

% Accurate evaluation of AGP is crucial because upscaled equations that forecast the gas output and lifespan of gas wells depend critically on the AGP at representative elementary volume scale. Therefore, using the current porous set-up, in figures \ref{perm_fixed_por} and \ref{perm_fixed_align}, we illustrated the    permeability correction factor (PCF), which is calculated as the ratio of apparent permeability ($\kappa$) to absolute permeability ($\kappa_\infty$), with varying alignment at fixed porosity and varying porosity and fixed alignment respectively.
Since up-scaled equations that predict the gas output and longevity of gas wells heavily rely on the AGP at representative elementary volume (REV) scale, it is essential to accurately evaluate AGP. For this reason, we first presented  the permeability correction factor (PCF) of different porous medium by varying the alignment at fixed porosity (Fig. \ref{perm_fixed_por}) and then varied  the porosity while keeping the alignment unchanged (Fig. \ref{perm_fixed_align}).
% 
% 
% in figures \ref{perm_fixed_por} and \ref{perm_fixed_align}, we presented  the permeability correction factor (PCF) of different porous medium by varying the alignment while keeping porosity fixed and vice-versa.
%, which is defined  as the ratio of apparent permeability ($\kappa$) to absolute permeability ($\kappa_\infty$), 
% of various alignments at fixed porosity and various porosity at fixed alignment, respectively, using the present porous setup. 
To calculate the absolute permeability ($\kappa_\infty$) for each set-up, we extrapolated the permeability calculated within the range from Kn = $10^{-3}$ to Kn = $10^{-1}$ and chose the value at Kn=$10^{-7}$ as $\kappa_\infty$.

  When the porosity is low ($\phi=0.75$) as shown in Fig. \ref{fixpor_75}, the setup with more non-uniformly distributed porous media shows a slightly higher value of PCF over the range of Kn. However, as the porosity is increased to higher values as one shown in Fig. \ref{fixpor_83} ($\phi=0.83$) and Fig. \ref{fixpor_90}($\phi=0.90$), the PCF starts to overlap on to each other for all the alignments. 
    Likewise, in Fig. \ref{perm_fixed_align}, we kept the alignment (uniformity/non-uniformity) of distributed porous media fixed and varied the porosity. 
%     In this case, we observed that AGP for low porosity starts to deviate towards a lower at higher values of Kn in all the alignmnet. Furthermore, as the alignmnet of the distribution of porous media is further incresed from $\theta_3$ to $\theta_6$, this deviation becomes more prominent.
    In this case, we observed that as Kn is increased (especially in the early transition regime) in all the alignments, the PCF for low porosity  has a higher value in comparison to the one with higher porosity. Additionally, this difference becomes more prominent as the dispersion of porous media is further increased from $\theta_3$ to $\theta_6$.
%     {\color{blue} 
    One reason for this behavior at the macroscale level could be that in less porous media the disturbance caused by one barrier propagates over a smaller distance and interacts with other obstacles before getting completely dispersed.
  The hydrodynamic disturbance may, however, tend to fade out or average out in highly porous media before encountering another barrier.
  This causes different kinds of fluid-solid interactions in addition to slip effect, which could change the magnitude of the velocity field. At the molecular level, large pores with high porosity have a lower rarefaction impact than smaller pores with low porosity because after colliding inside one pore's throat, the gas molecule must travel a greater distance before colliding with another obstruction in highly porous medium as compared one with small porosity.
%   } 
%     This leads us to conclude that, in the current porous setup, the non-uniformity (alignment) of porous material, which is one of the factor in determing the tortuosity of the material, has less of an impact on PCF than does porosity.
Overall, figures \ref{perm_fixed_por} and \ref{perm_fixed_align} led us to the conclusion that, in the current porous setting, porosity has a greater impact on PCF than does non-uniformity (alignment) of porous material, which is one of the factors in determining the tortuosity of the material.
  \begin{figure}
      \includegraphics[height=60mm,width=100mm]{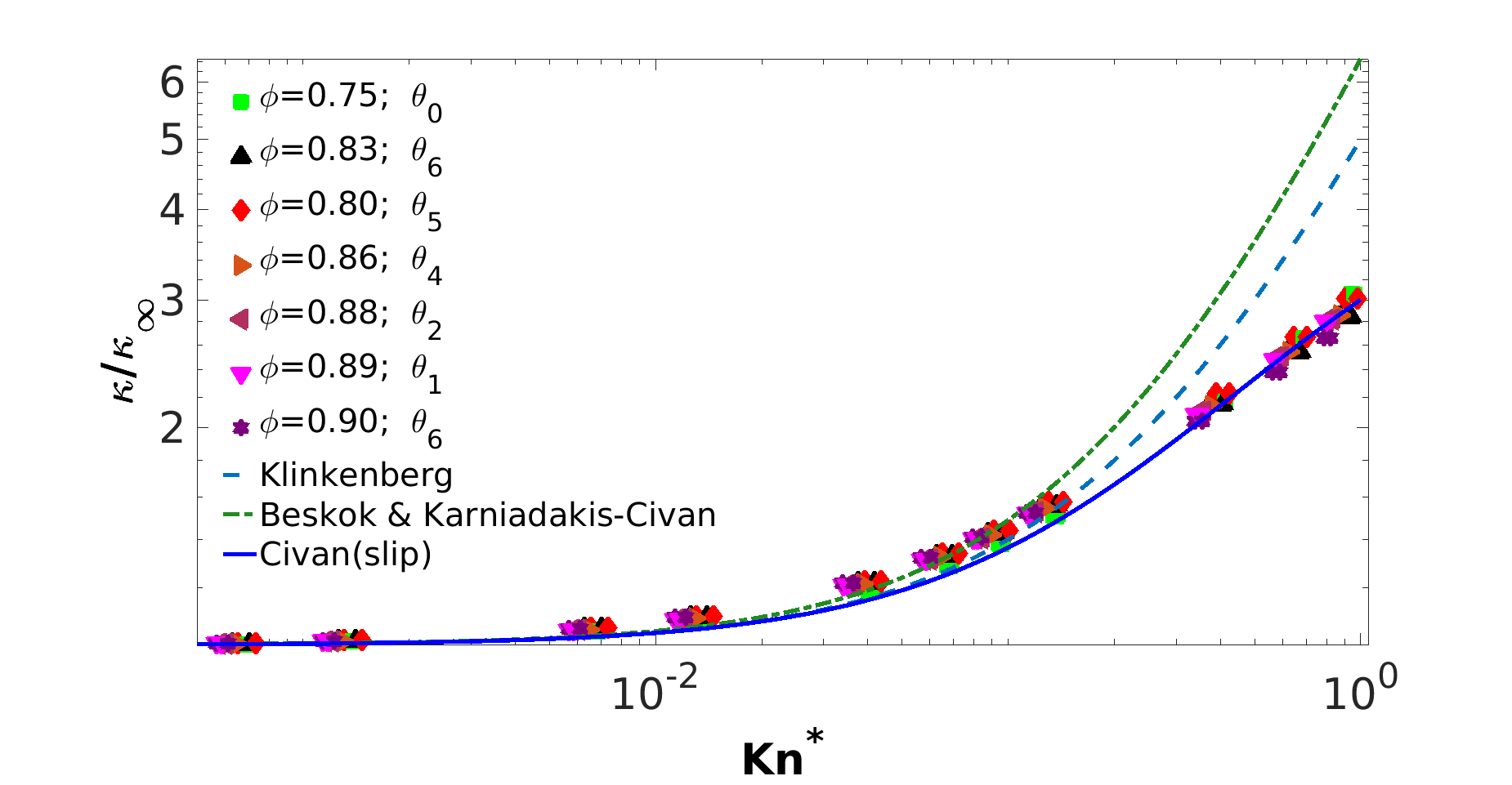}
      \caption{The permeability correction factor (PCF) as a function of an effective ${\rm Kn}^\star$ at randomly selected porous media configuration. The first-order correlation by Klinkenberg (Eq. \eqref{klinkenberg}) \cite{klinkenberg1941permeability}, the  second-order correlation given by Beskok \& Karniadakis-Civan (Eq. \eqref{beskok_karniadakis})\cite{beskok1999report,civan2010effective} and the correlation in the slip flow regime suggested by Civan (Eq. \eqref{civan-slip})\cite{civan2010effective} are  also presented as a reference.}
      \label{mixed_perm}
  \end{figure}

%   According to recent studies, if the Kn is defined by the effective pore size, which is determined using the proper porosity and tortusity, the PCF for different porosity and pore structures is found to collapse to one curve. It can also be accomplished by scaling Kn with a suitable function of porosity () and tortuisty (), as recommended by Wu et al. \cite{}. 
  In recent studies, the direct simulation Monte Carlo (DSMC) approach was employed to demonstrate that even for complicated porous structures, the apparent permeability of a porous media can be predicted through the Klinkenberg correlations using  fundamental and measurable descriptors of the pore structure \cite{wu2016non,wu2017apparent,yang2018investigation,su2020gsis}. 
  %This was achieved by using an effective Kn which is  determined by properly porosity and tortuosity.
%   
% %   According to recent studies, if the Kn is defined by the effective pore size, which is determined using the properly porosity and tortusity, 
% It was observed that the PCF for different porosity and pore structures is found to collapse to one curve. It can also be accomplished by scaling Kn with a suitable function of porosity () and tortuisty (), as recommended by Wu et al. \cite{}.
This can be accomplished by scaling Kn with an appropriate function of porosity and tortuosity, as suggested by Wu et al. \cite{wu2017apparent}, or by using an effective pore size that is determined by porosity and tortuosity \cite{yang2018investigation}, as it was noticed that using such arguments, the PCF for different porosity and pore structures is found to collapse onto one curve.
In a similar manner, we identified an empirical function, $g=\phi/T^2$, to determine an effective Knudsen number ($\rm Kn^\star$) for the current collection of porous media, given as ${\rm Kn}^\star={\rm Kn}/{g}$. 
This can be inferred indirectly as  rescaling of the pore size using parameters related to pore structure, porosity, and pore scale flow, hydraulic tortuosity (T).
% This can  indirectly be interpreted as re-scaling the  pore-size with parameter associated with pore-structure like porosity and with pore-scale flow like hydraulic tortuosity (T). }
%   In this spirit, after some investigation, we discovered an emperical function, $g=\phi/T^2$, to establish an effective Knudsen number ($\rm Kn^\star$) for the current collection of porous media, as $ {\rm Kn}^\star={\rm Kn}/{g}.$  
  %In this spirit, after some analysis, we found an emperical function $g(\phi,T)$ as $g=\phi/T^2$ to define an effective Knudsen number (${\rm Kn}^\star$), for the present set of porous media, as
%   \begin{equation}
%   
%   \end{equation}
  In Fig. \ref{mixed_perm}, the PCF is plotted with respect to the effective Knudsen number ${\rm Kn}^\star$. The figure shows that for all the geometries (chosen at random), the PCF lie almost entirely along a single line.
  %{\color{blue}
 Figure \ref{mixed_perm} further demonstrates that the PCF and the three proposed correlations- Klinkenberg (Eq. \eqref{klinkenberg}) \cite{klinkenberg1941permeability}, Beskok \& Karniadakis-Civan (Eq. \eqref{beskok_karniadakis})\cite{beskok1999report,civan2010effective}, and Civan (Eq. \eqref{civan-slip})\cite{civan2010effective}-are in fair agreement up until the slip flow regime. However, in the early transition regime, the best agreement  is with Civan \cite{civan2010effective}, which was initially suggested for the flow close to slip regime. Figure 3 illustrates that a generalized correlation can be obtained up until the early transition regime with an appropriate effective Knudsen number.
  %}

  \section{\label{conclude}Conclusion and Outlook}
%   We investigated the pressure-driven gas transport characteristics across a homogeneously arranged porous medium containing micro- and nanopores for moderate to high porosities of $0.75 <\phi< 0.90$ for various Knudsen number using the extended lattice Boltzmann method.
  We investigated the  gas transport characteristics flowing across porous medium containing micro- and nanopores for moderate to high porosities of $0.75 <\phi< 0.90$ from the continuum to early-transition regimes in a systematic manner using the extended lattice Boltzmann method with a focus on two physical parameters of the flow namely, tortuosity and permeability. {\it A priori}, it is not obvious that tortuosity and porosity in general have a universal relationship. However, such correlation can arise, at least for some porous material types.  With this aim, we explored the gas transport in homogeneously arranged porous media which showed an empirical power-law behaviour between the two as $T-1\sim (1-\phi)^\gamma$. An further investigation across various Kn shows that,  in the slip flow regime, the Kn-dependency alters the exponent of the power as $T-1\sim (1-\phi)^{\gamma+f_n}$ with $f^n\sim{\rm Kn^{0.6}}$, giving a generalized relation  between the two.

   In addition, we discovered that permeability correction factor (PCF) with respect to Kn, which appears as a result of rarefaction, fall nearly on one single line up until early transition regime, indicating a generalized correlation, with appropriate scaling of Kn with parameters like porosity and hydraulic tortuosity.  Moreover, how the Kn should be properly scaled using $\phi$ and $T$ in a heterogeneous and tortuous environment may  also be  crucial to examine other phenomena such as the dispersive  transport of a scalar through complex media with pores size ranging from micro to nano-meters \cite{meigel2022dispersive}.
It should be emphasized that the implementation of the diffuse wall boundary condition, which requires wall-normals, was simple due to the simplicity of the current 2D set-up formed by arranging circular obstacles in a homogeneous pattern at different orientations. The distribution of pores in a real rock, however, is highly heterogeneous and tortuous. In future, we will  explore the fluid flow in a realistic low-porosity geometry of natural rock imaged by a multislice micro-CT scanner\cite{jiang2014changes,jiang2023upscaling} by  utilizing a boundary condition that was more practical for such a scenario and does not require calculating wall normals as proposed in Ref. \cite{singh2017impact}.

\section{Acknowledgment}
  S.S. acknowledges the financial support by the Leverhulme Early Career Fellowship (ECF-2019-100). S.S. also acknowledges the use of the Scientific Computing Research Technology Platform and associated support services at the University of Warwick, in the completion of this work.
 
\appendix
\section{Lattice Boltzmann Method}
\label{LBM}
%   The lattice Boltzmann method has emerged as an effective alternate to Navier Stokes solver over the past few decades \cite{benzi1992lattice,succi2001lattice}.  
  The conventional LBM framework is based on the Boltzmann equation with single relaxation approximation also known as (BGK) approximation \cite{bhatnagar1954model}.
  The discrete form of such a equation requires a set of discrete populations $f=\{f_i\}$ and corresponding to which there exists a pre-defined discrete velocities ${\pmb c}_i$ $(i=1,\cdots,N)$  \cite{succi2001lattice,benzi1992lattice} and has the following form
  \begin{equation}
 f_{i}({\pmb x}+{\pmb c}\Delta t, t+\Delta t)=f_{i}({\pmb x}, t)+\Omega_{i}(f)+ \Delta t { F}_i.
 \label{dis_BGK}
\end{equation}
Here, $F_i$ corresponds to $i$-th component of external force and the BGK collision approximation, $\Omega_{i}(f)$,  given as
\begin{equation}
 \Omega_{i}(f)=\frac{\Delta t}{\tau}(f_{i}^{\rm eq}-f_{i}({\pmb x},t)],
\end{equation}
  dictates the relaxation of  distribution function to an equilibrium Maxwell-Boltzmann function, $f^{\rm eq}$, as given in Eq. \eqref{dis_eq},  
 at the rate of $\tau^{-1}$.
%The discrete version of the  Maxwell-Boltzmann  used in   till second order  in  Mach number (Ma) is
 \begin{align}
 %\label{Max_boltz}
\begin{split}
\label{dis_eq}
{f}_i^{\rm eq} &= w_i\rho\Biggl[1 +
\frac{ {\pmb c}_{i}\cdot{\pmb u}}{ \, c_s^2} 
+ \frac{({\pmb c}_{i}\cdot{\pmb u})^2}{2\, \,   c_s^4} 
-\frac{({\pmb u}\cdot{\pmb u})}{2\, \,   c_s^2} 
  \Biggr].
 \end{split}
\end{align}
The two-dimensional model (D2Q9) which is chosen for the present study has the following nine discrete velocities
\begin{equation}
{\pmb c}_i = \begin{cases} (0,0) &\mbox{if } i = 0 \\
 \left(\cos{\frac{(i-1)\pi}{4}},\sin{\frac{(i-1)\pi}{4}}\right)& \mbox{if } i=1,2,3,4\\
 \sqrt{2}\left(\cos{\frac{(i-1)\pi}{4}},\sin{\frac{(i-1)\pi}{4}}\right) &\mbox{if } i =5,6,7,8,
 \end{cases}  
\end{equation}
with the corresponding weights as
\begin{equation}
w_i = \begin{cases} \frac{4}{9} &\mbox{for } i = 0 \\
  \frac{1}{9}& \mbox{for } i=1,2,3,4\\
 \frac{1}{36} &\mbox{for } i =5,6,7,8.
 \end{cases}  
\end{equation}
The lattice sound speed $ c_s$  that appeared in the Eq. \eqref{dis_eq} is related to the magnitude of discrete velocity as  $c^2=3c_s^2$.
The relevant hydrodynamic macroscopic moments, like density ($\rho$), momentum density $(\rho {\pmb u})$ and momentum flux (${\pmb \Pi}$)
can be obtained by linear weighted sums as 
$ \rho({\pmb x},t)=\sum_i f_i, \quad
 \rho({\pmb x},t) {\pmb u}({\pmb x},t)=\sum_i f_i {\pmb c}_i,\quad \textrm{and} \quad
 {\pmb \Pi}({\pmb x},t)=\sum_i f_i({\pmb c}_i{\pmb c}_i-c_s^2{\pmb \delta}),
 $
 respectively with $ {\pmb \delta}$ being the identity matrix.
 Finally, the term $F_i$ in Eq. \eqref{dis_BGK}, which is the $i^{\rm th}$ component of the body force, is given as \cite{guo2002discrete}:
\begin{equation}
 F_i=w_i\rho\left[\frac{{\pmb g}\cdot{\pmb c}_i}{c_s^2}+\frac{({\pmb g}{\pmb u}+{\pmb u}{\pmb g})}{2c_s^2}{\pmb :}({\pmb c}_i{\pmb c}_i-c_s^2 {\pmb \delta})\right],
\end{equation}
where ${\pmb g}$ is the constant acceleration vector. In the subsequent section, we will briefly discuss a regularization scheme which was introduced to filter out the nonphysical effect from finite Kn flow.
% 
%   {\bf write about the Regularization and KB}
%   {\color{blue}  The physical meaning of the regularization is quite trans-
% parent: It filters out the nonhydrodynamic moments from
% the postcollision distribution function, so as to minimize
% their unphysical effect on the macroscopic behavior of the
% flow}
    \subsection{Regularization Scheme}
    Initially proposed to resolve the issue of stability of  high viscous flows, the regularization process, as introduced by Chen and co-workers \cite{zhou2006simulation,zhang2006efficient} and Latt and Chopard \cite{latt2006lattice}, turns out be one of the major ingredient for the finite Kn flow \cite{montessori2015lattice} in the lattice Boltzmann framework. In regularization process, non-hydrodynamic (ghost) modes are filtered out by dividing the post streaming distribution function into two parts as:
    \begin{equation}
   f_i=f_{i}^{\rm eq}+f_{i}^{\rm neq}.
  \end{equation}
  The information about the hydrodynamic modes is contained in $f_{i}^{\rm eq}$  and to remove the information of non-hydrodynamic modes from $f_{i}^{\rm neq}$, it is converted into a new distribution function $f^{\rm Reg}_{i}$ which is then defined in terms of hydrodynamic moments $(\rho,{\pmb u},{\pmb \Pi})$ and has the following discrete form:
\begin{equation}
 f^{\rm Reg}_{i}=\frac{w_i}{2 c_s^4}({\pmb c}_i{\pmb c}_i- c_s^2 {\pmb \delta}):{\pmb \Pi}^{\rm neq},
\end{equation}
where ${\pmb \Pi}^{\rm neq}$ is the non-equilibrium part of momentum flux and is given as ${\pmb \Pi}^{\rm neq}=\sum_i f^{\rm neq}({\pmb c}_i{\pmb c}_i- c_s^2 {\pmb \delta})$. The  streaming step of lattice Boltzmann method takes the following form after the regularization process:  
\begin{equation}
 f_{i}({\pmb x}+{\pmb c}\Delta t, t+\Delta t)=f_{i}^{\rm eq}+
 \left(1-\frac{\Delta t}{\tau}\right)f_{i}^{\rm Reg} + \Delta t { F}_i.
 \label{dis_BGK_REG}
\end{equation}   
    \subsection{Kinetic Boundary Condition}
    The no-slip boundary conditions based on bounce back mechanism, where   the directions
of the incoming distribution functions are simply reversed when
it encounters the boundary node, as the name suggests fails as to uncover the slip velocity at the boundary at finite Kn. To overcome this shortcoming of the method, a  diffusively reflecting solid wall   boundary condition was introduced by Ansumali and Karlin \cite{ansumali2002kinetic}. Based on kinetic theory interpretation, this boundary condition redistributes the population coming towards the wall in such a way that  mass-balance and normal-flux
conditions are fulfilled. In the discrete sense, the distribution function at the boundary/wall takes the following form
\begin{equation}
f_i({\pmb x}_w,t)= K f_{i}^{\rm eq}(\rho, {\pmb u}_w),
\label{diffuse_bc}
\end{equation}
where subscript $w$ denotes the wall and
\begin{equation}
K= \frac{\sum_{{\pmb c}_i\cdot {\pmb n}<0}|({\pmb c}_i-{\pmb u}_w)\cdot {\pmb n}| f_i}{\sum_{{\pmb c}_i\cdot {\pmb n}>0}|({\pmb c}_i-{\pmb u}_w)\cdot {\pmb n}|f_{i}^{\rm eq}(\rho, {\pmb u}_w)},
\end{equation}
with ${\pmb n}$ being the unit normal direction. 
\begin{figure}[h]
\subfigure[$n_x$]{
\label{nX}
   \includegraphics[scale=0.125]{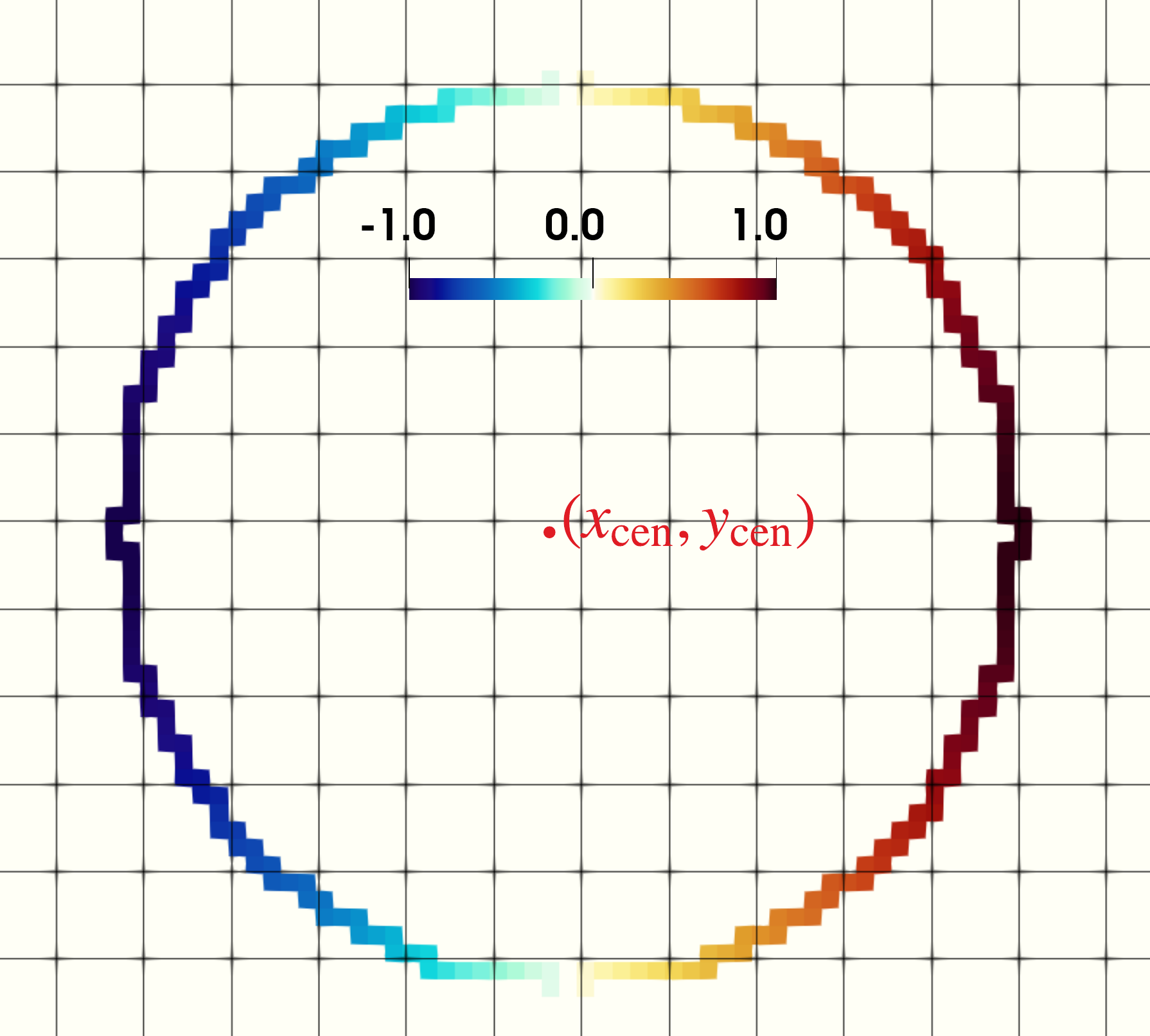}}
\subfigure[$n_y$]{
\label{nY}
\includegraphics[scale=0.125]{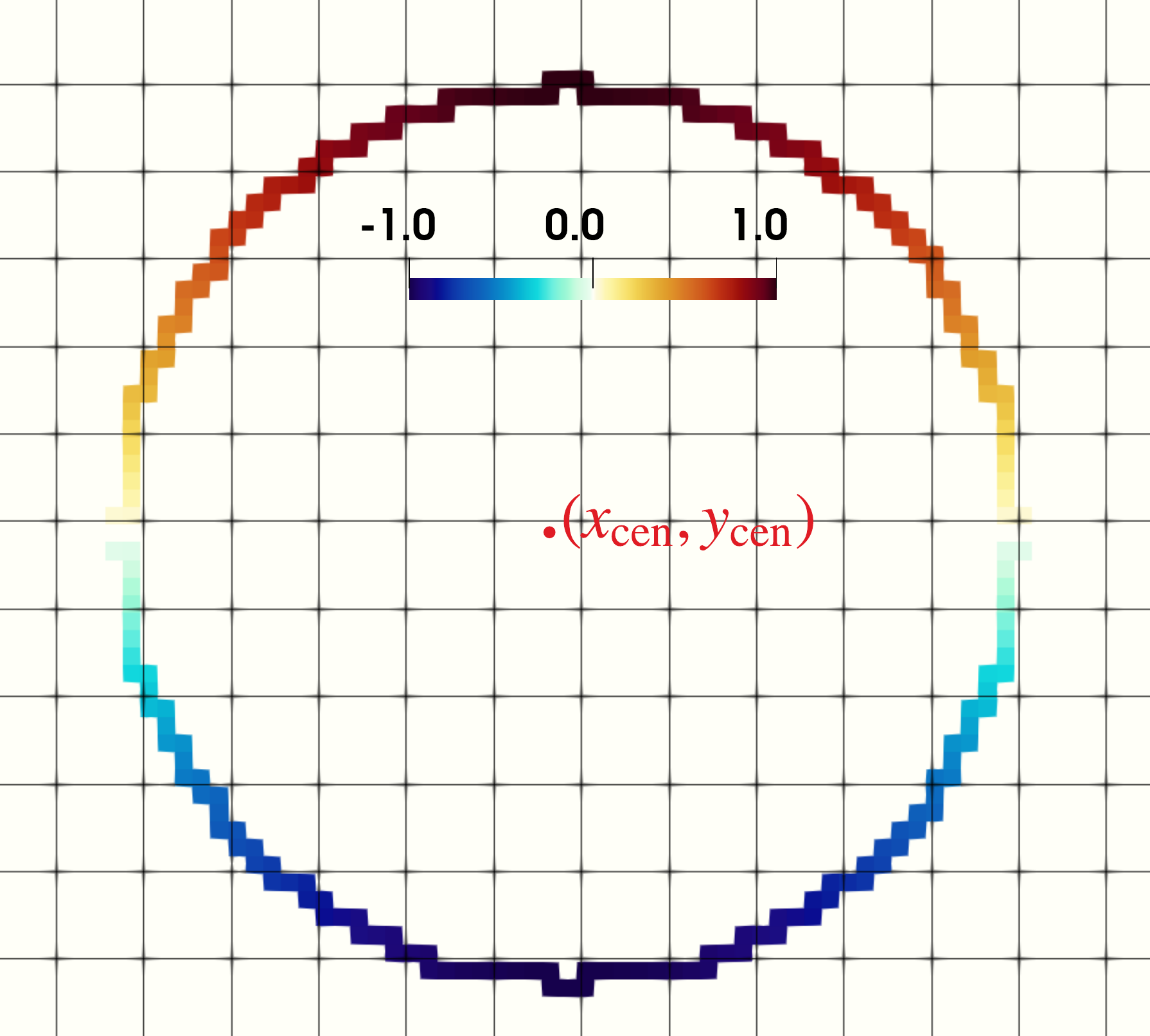}}
\caption{Unit normals at the solid boundaries}
  \end{figure}
   The term K can be
understood as the ratio of  outgoing flux from the wall
and  incoming equilibrium flux coming towards the wall. When the boundaries are stationary (${\pmb u}_w=0$), the term K can be written as:
\begin{equation}
\label{knorm}
K= \frac{\sum_{A_i<0}|A_i| f_i}{\sum_{A_i>0}|A_i|f_{i}^{\rm eq} },
\end{equation}
 with $A_i={c}_{ix}{ n}_{x}+{c}_{iy}{ n}_{y}$. To implement this boundary condition
one needs the information of unit normals at the boundaries. Fig \ref{nX} and \ref{nX} respectively represents the  unit normals in the $x$-direction, $n_x$, and the same in $y$-direction, $n_y$ \cite{singh2017impact,singh2017influence} for a circular obstacle. In the present study, the normals for an individual circle with center $(x_{\rm cen},y_{\rm cen})$, are calculated as  
\begin{equation}
\hat{\pmb n}=\frac{{\pmb x}-{\pmb x}_{\rm cen}}{|{\pmb x}-{\pmb x}_{\rm cen}|}.
\end{equation}
% {\color{blue}
% 
% After the streaming step, a further regularization step, which consists in filtering out
% high-order non-hydrodynamic nodes that may emerge,
% }
%     {\color{red}
%     Recognizing this shortcoming of LBGK for small-scale
% flow and exploiting the kinetic origin of LBM, a diffusively
% reflecting solid wall boundary condition was proposed by
% Ansumali and Karlin [16]. The basic idea behind this boundary
% condition is to redistribute the populations coming towards
% the wall such that it follows mass-balance and normal-flux
% conditions. The discrete version of the boundary condition is
% with n being the unit normal direction. The term K can be
% understood as the ratio of total outgoing flux to the wall
% and total incoming equilibrium flux from the wall. This
% interpretation of the term K will be utilized later in Sec. IV
% to construct a suitable boundary condition for porous media.}

 \bibliographystyle{unsrt}
 \bibliography{references}
\end{document}